\documentclass{appolb}%
\usepackage{amssymb}
\usepackage{amsmath}
\usepackage{epsfig}
\usepackage{amsfonts}
\usepackage{graphicx}%
\begin{document}

\title{Interaction of the pseudoscalar glueball with (pseudo)scalar mesons and
nucleons\thanks{Presented by Walaa Eshraim at the Workshop \textquotedblleft Excited QCD
2012\textquotedblright\ , 6-12 May 2012, Peniche (Portugal)}}
\author{Walaa. I. Eshraim, Stanislaus Janowski, Antje Peters, Klaus Neuschwander, and
Francesco Giacosa
\address{Institute for Theoretical Physics, Johann Wolfgang Goethe University,
Max-von-Laue-Str.\ 1, D--60438 Frankfurt am Main, Germany} }
\maketitle

\begin{abstract}
We study the interactions of the pseudoscalar glueball with scalar and
pseudoscalar quark-antiquark meson fields and with the nucleon and its chiral
partner. In both cases we introduce the corresponding chiral Lagrangian and
discuss its properties. We calculate the mesonic and baryonic decays of a
pseudoscalar glueball with mass of about 2.6 GeV as predicted by Lattice simulations.

\end{abstract}



\section{Introduction}

The investigation of the properties of bound state of gluons, the so-called
glueballs, represents an important step toward the understanding of the
nonperturbative aspects of Quantum Chromodynamics (QCD). The search for
glueballs is also relevant in the framework of hadron phenomenology, as they
might explain the nature of some enigmatic mesonic resonances (see Ref.
\cite{review} and refs. therein).

Lattice QCD is a well-established non-perturbative approach to solve QCD:
within this context the glueball spectrum has been obtained \cite{Morningstar}%
, where the lightest glueball has $J^{PC}=0^{++}$ quantum numbers and a mass
of about 1.6 GeV. This energy region has been studied in a variety of
effective approaches, e.g. Refs. \cite{scalars,stani}. The second lightest
glueball has been predicted to be a tensor ($J^{PC}=2^{++}$), see also Ref.
\cite{tensor} for a related phenomenological discussion. The third lightest
state is a pseudoscalar glueball ($J^{PC}=0^{-+}$) with a mass of about 2.6
GeV. This value represents the starting point of our investigation of the
properties of the pseudoscalar glueball \cite{our} (for scenarios with a lower
mass see Ref. \cite{mixetas} and refs. therein).

Namely, we study the interactions of the pseudoscalar glueball, denoted as
$\tilde{G}$, to scalar and pseudoscalar mesons: we discuss the symmetry
properties of the effective Lagrangian introduced in Ref. \cite{our} and we
present the results for the branching ratios for the two-body decays (one
scalar and one pseudoscalar state) and for the three-body decays (three
pseudoscalar states). Then, we comment on a particular interference problem,
only mentioned in\ Ref. \cite{our}, which emerges from the subsequent decay of
a scalar meson of the two-body decay into two pseudoscalar states: both decay
mechanism end up in the same final states and therefore care is needed. Next,
we describe (to our knowledge for the first time) the interaction of
$\tilde{G}$ with baryons: we introduce the chiral effective Lagrangian which
couples $\tilde{G}$ to the nucleon field and its chiral partner. This
Lagrangian describes also the proton-antiproton conversion process $\bar
{p}p\rightarrow\tilde{G}$, which can take place in the planned PANDA
experiment at the upcoming FAIR facility in Darmstadt \cite{panda}, in which
the (center of mass) energy range above 2.5 GeV will be investigated.

\section{Interaction with (pseudo)scalar mesons}

The effective Lagrangian which couples the pseudoscalar glueball field,
$\tilde{G}$ with quantum numbers $J^{PC}=0^{-+}$ to scalar and pseudoscalar
mesons read \cite{our,schechter}
\begin{equation}
\mathcal{L}_{\tilde{G}\text{-mesons}}^{int}=ic_{\tilde{G}\Phi}\tilde{G}\left(
\text{\textrm{det}}\Phi-\text{\textrm{det}}\Phi^{\dag}\right)  \text{ ,}
\label{intlag}%
\end{equation}
where $c_{\tilde{G}\phi}$ is the (unknown) coupling constant. The scalar and
pseudoscalar mesons are organized in the multiplet $\Phi$ \cite{dick}:%
\begin{equation}
\Phi=\frac{1}{\sqrt{2}}\left(
\begin{array}
[c]{ccc}%
\frac{(\sigma_{N}+a_{0}^{0})+i(\eta_{N}+\pi^{0})}{\sqrt{2}} & a_{0}^{+}%
+i\pi^{+} & K_{S}^{+}+iK^{+}\\
a_{0}^{-}+i\pi^{-} & \frac{(\sigma_{N}-a_{0}^{0})+i(\eta_{N}-\pi^{0})}%
{\sqrt{2}} & K_{S}^{0}+iK^{0}\\
K_{S}^{-}+iK^{-} & \bar{K}_{S}^{0}+i\bar{K}^{0} & \sigma_{S}+i\eta_{S}%
\end{array}
\right)  \text{ } \label{phimatex}%
\end{equation}
which transforms as $\Phi\rightarrow U_{L}\Phi U_{R}^{\dagger}$ under chiral
transformations of the group $U(3)_{R}\times U(3)_{L}$, whereas $U_{L(R)}%
=e^{-i\theta_{L(R)}^{a}t^{a}}$ is an element of $U(3)_{R(L)}.$ The
pseudoscalar glueball $\tilde{G}$ consists of gluons and is a chirally
invariant object. It follows that the Lagrangian (\ref{intlag}) is invariant
under $SU(3)_{R}\times SU(3)_{L}$ transformations, but is not invariant under
the axial $U_{A}(1)$ transformation, because:
\[
\mathrm{det}\Phi\rightarrow\mathrm{det}U_{A}\Phi U_{A}=e^{-i\theta_{A}%
^{0}\sqrt{2N_{f}}}\mathrm{det}\Phi\neq\mathrm{det}\Phi\text{ .}%
\]

We now turn to discrete symmetries. The parity transformation $\mathcal{P}$ of
the multiplet $\Phi$ reads $\Phi(t,\vec{x})\rightarrow\Phi^{\dagger}%
(t,-\vec{x})$ and that of the glueball reads $\tilde{G}(t,\vec{x}%
)\rightarrow-\tilde{G}(t,-\vec{x}).$ It is then easy to verify that the
Lagrangian (\ref{intlag}) is parity invariant. Under charge conjugation
$\mathcal{C}$ the transformations $\Phi\rightarrow\Phi^{T}$ and $\tilde
{G}\rightarrow\tilde{G}$ hold, in virtue of which the Lagrangian
(\ref{intlag}) is also left invariant.

The assignment of the quark-antiquark fields in our work is as follows: (i) In
the pseudoscalar sector the fields $\vec{\pi}$ and $K$ represent the pions or
the kaons, respectively. The bare fields $\eta_{N}\equiv\left\vert \bar
{u}u+\bar{d}d\right\rangle /\sqrt{2}$ and $\eta_{S}\equiv\left\vert \bar
{s}s\right\rangle $ are the non-strange and strange mixing contributions of
the physical states $\eta$ and $\eta^{\prime}$ . (ii) In the scalar sector we
assign the field $\vec{a}_{0}$ to the physical isotriplet state $a_{0}(1450)$
and the scalar kaon fields $K_{S}$ to the resonance $K_{0}^{\star}(1430).$ The
fields $\sigma_{N}\equiv\left\vert \bar{u}u+\bar{d}d\right\rangle /\sqrt{2}$
and $\sigma_{S}\equiv\left\vert \bar{s}s\right\rangle $ correspond to the
physical resonances \thinspace$f_{0}(1370)$ and $f_{0}(1710)$. The small
mixing of the bare fields $\sigma_{N}$ and $\sigma_{S}$ is neglected here
\cite{dick}.

To evaluate the decays of the pseudoscalar glueball $\tilde{G}$ we have to
take into account that the spontaneous breaking of chiral symmetry takes
place, which implies the shift of the scalar-isoscalar fields as $\sigma
_{N}\rightarrow\sigma_{N}+\phi_{N}$ and $\sigma_{S}\rightarrow\sigma_{S}%
+\phi_{S}$, where $\phi_{N}$ and $\phi_{S}$ represent the chiral non-strange
and strange condensates. In addition, due to the fact that also (axial-)vector
mesons are present in the full Lagrangian \cite{stani,dick,denis}, one has
also to `shift' the axial-vector fields and to redefine the renormalization
constant of the pseudoscalar fields, $\vec{\pi}\rightarrow Z_{\pi}\vec{\pi}$ ,
$K\rightarrow Z_{K}K$, $\eta_{N,S}\rightarrow Z_{\eta_{N,S}}\eta_{N,S}$, where
the quantities $Z_{i}$ are the wave function renormalization constants. The
theoretical results for the two-body and three-body branching ratios of the
pseudoscalar glueball $\tilde{G}$ as evaluated from Eq. (\ref{intlag}) are
summarized in Table I.a and I.b for the mass $M_{\tilde{G}}=2.6$ GeV, see also
Ref. \cite{our}. Note, the ratios are independent on the unknown coupling
$c_{\tilde{G}\Phi}$ and represent a prediction of our approach.

\begin{center}
\bigskip%
\begin{tabular}
[c]{|c|c|}\hline
Quantity & Value\\\hline
$\Gamma_{\tilde{G}\rightarrow KK\eta}/\Gamma_{\tilde{G}}^{tot}$ &
$0.049$\\\hline
$\Gamma_{\tilde{G}\rightarrow KK\eta^{\prime}}/\Gamma_{\tilde{G}}^{tot}$ &
$0.019$\\\hline
$\Gamma_{\tilde{G}\rightarrow\eta\eta\eta}/\Gamma_{\tilde{G}}^{tot}$ &
$0.016$\\\hline
$\Gamma_{\tilde{G}\rightarrow\eta\eta\eta^{\prime}}/\Gamma_{\tilde{G}}^{tot}$
& $0.0017$\\\hline
$\Gamma_{\tilde{G}\rightarrow\eta\eta^{\prime}\eta^{\prime}}/\Gamma_{\tilde
{G}}^{tot}$ & $0.00013$\\\hline
$\Gamma_{\tilde{G}\rightarrow KK\pi}/\Gamma_{\tilde{G}}^{tot}$ &
$0.46$\\\hline
$\Gamma_{\tilde{G}\rightarrow\eta\pi\pi}/\Gamma_{\tilde{G}}^{tot}$ &
$0.16$\\\hline
$\Gamma_{\tilde{G}\rightarrow\eta^{\prime}\pi\pi}/\Gamma_{\tilde{G}}^{tot}$ &
$0.094$\\\hline
\end{tabular}
$%
\begin{array}
[c]{c}%
\\
\\
\end{array}
$%
\begin{tabular}
[c]{|c|c|}\hline
Quantity & Value\\\hline
$\Gamma_{\tilde{G}\rightarrow KK_{S}}/\Gamma_{\tilde{G}}^{tot}$ &
$0.059$\\\hline
$\Gamma_{\tilde{G}\rightarrow a_{0}\pi}/\Gamma_{\tilde{G}}^{tot}$ &
$0.083$\\\hline
$\Gamma_{\tilde{G}\rightarrow\eta\sigma_{N}}/\Gamma_{\tilde{G}}^{tot}$ &
$0.028$\\\hline
$\Gamma_{\tilde{G}\rightarrow\eta\sigma_{S}}/\Gamma_{\tilde{G}}^{tot}$ &
$0.012$\\\hline
$\Gamma_{\tilde{G}\rightarrow\eta^{\prime}\sigma_{N}}/\Gamma_{\tilde{G}}%
^{tot}$ & $0.019$\\\hline
\end{tabular}

Table I.a (left): Branching ratios for the three-body decays $\tilde
{G}\rightarrow PPP$.

Table I.b (right): Branching ratios for the two-body decays $\tilde
{G}\rightarrow SP$.
\end{center}

An interesting and subtle issue is the following: the scalar states decay
further into two pseudoscalar ones. For instance, $K_{S}\equiv$ $K_{0}^{\ast
}(1430)$ decays into $K\pi$. There are then two possible decay amplitudes for
the process $\tilde{G}\rightarrow KK\pi$: one is the direct decay mechanism
reported in Table I.a, the other is the decay chain $\tilde{G}\rightarrow
KK_{S}\rightarrow KK\pi$. The immediate question is, if interference effects
emerge which spoil the results presented in Table I.a and I.b. Namely, simply
performing the sum of the direct three-body decay (Table I.a) and the
corresponding two-body decay (table I.b) is not correct.

We now describe this point in more detail using the neutral channel $\tilde
{G}\rightarrow K^{0}\bar{K}^{0}\pi$ as an illustrative case. To this end, we
describe the coupling $K_{S}=K_{0}^{\ast}$ to $K\pi$ via the Lagrangian
\begin{equation}
\mathcal{L}_{K_{S}K\pi}=gK_{0}^{\ast}\bar{K}_{0}\pi^{0}+\sqrt{2}gK_{0}^{\ast
}K^{-}\pi^{+}+h.c.\text{ .}%
\end{equation}
The coupling constant $g=2.73$ GeV is obtained by using the experimental value
for the total decay width $\Gamma_{K_{0}^{\ast}}=270$ MeV \cite{pdg}. The full
amplitude for the process $\tilde{G}\rightarrow K^{0}\bar{K}^{0}\pi^{0}$
results as the sum
\begin{equation}
\mathcal{M}_{\tilde{G}\rightarrow K^{0}\bar{K}^{0}\pi^{0}}^{\text{full}%
}=\mathcal{M}_{\tilde{G}\rightarrow K^{0}\bar{K}^{0}\pi^{0}}^{\text{direct}%
}+\mathcal{M}_{\tilde{G}\rightarrow\bar{K}^{0}K_{S}^{0}.\rightarrow K^{0}%
\bar{K}^{0}\pi^{0}}^{\text{via}K_{S}}+\mathcal{M}_{\tilde{G}\rightarrow
K^{0}\bar{K}_{S}^{0}.\rightarrow K^{0}\bar{K}^{0}\pi^{0}}^{\text{via}\bar
{K}_{S}}\text{ }%
\end{equation}
Thus for the decay width we obtain
\begin{align}
\Gamma_{\tilde{G}\rightarrow K^{0}\bar{K}^{0}\pi^{0}}^{\text{full}} &
=\Gamma_{\tilde{G}\rightarrow K^{0}\bar{K}^{0}\pi^{0}}^{\text{direct}}%
+\Gamma_{\tilde{G}\rightarrow K^{0}K_{S}^{0}\rightarrow K^{0}\bar{K}^{0}%
\pi^{0}}^{\text{via}K_{S}}+\nonumber\\
&  \Gamma_{\tilde{G}\rightarrow K^{0}\bar{K}_{S}^{0}.\rightarrow K^{0}\bar
{K}^{0}\pi^{0}}^{\text{via}\bar{K}_{S}}+\Gamma_{\tilde{G}\rightarrow K^{0}%
\bar{K}^{0}\pi^{0}}^{\text{mix}}%
\end{align}
where $\Gamma_{\tilde{G}\rightarrow K^{0}\bar{K}^{0}\pi^{0}}^{\text{mix}}$ is
the sum of all interference terms. We can then investigate how large the
mixing term $\Gamma_{\text{mix}}$ is, and thus the error done in neglecting
it. The explicit calculation for the $K^{0}\bar{K}^{0}\pi^{0}$ case gives a
relative error of%
\begin{equation}
\left\vert \frac{\Gamma_{\tilde{G}\rightarrow K^{0}\bar{K}^{0}\pi^{0}%
}^{\text{mix}}}{\Gamma_{\tilde{G}\rightarrow K^{0}\bar{K}^{0}\pi^{0}%
}^{\text{direct}}+\Gamma_{\tilde{G}\rightarrow K^{0}K_{S}^{0}\rightarrow
K^{0}\bar{K}^{0}\pi^{0}}^{\text{via}K_{S}}+\Gamma_{\tilde{G}\rightarrow
K^{0}\bar{K}_{S}^{0}.\rightarrow K^{0}\bar{K}^{0}\pi^{0}}^{\text{via}\bar
{K}_{S}}}\right\vert \approx%
\begin{array}
[c]{c}%
7.3\text{ \% (}g>0\text{)}\\
2.2\text{ \% (}g<0\text{)}%
\end{array}
\text{ }%
\end{equation}
Present results from the model in Ref. \cite{dick} show that $g<0$: the
estimates presented in Ref. \cite{our} can be regarded as upper limits. We
thus conclude that the total error for the channel $\tilde{G}\rightarrow
K^{0}\bar{K}^{0}\pi^{0}$ is not large and can be neglected at this stage.
However, in future more detailed and precise theoretical predictions, these
interference effects should also be taken into account.

\section{Interaction with baryons}

In the planned PANDA experiment at FAIR \cite{panda} antiprotons collide on a
proton rich target. It is then also interesting to study how the pseudoscalar
glueball interacts with the nucleon (and with its chiral partner). In the
so-called mirror assignment \cite{detar,gallas}, one starts from two nucleon
fields $\Psi_{1}$ and $\Psi_{2}$ which transform in under chiral
transformations as follows:
\begin{equation}
\Psi_{1R(L)}\longrightarrow U_{R(L)}\Psi_{1R(L)}\text{ , }\Psi_{2R(L)}%
\longrightarrow U_{L(R)}\Psi_{2R(L)}\text{ .}%
\end{equation}
In this way it is possible to write down a chirally invariant mass term of the
type
\begin{equation}
\mathcal{L}_{m_{0}}=-m_{0}\left(  \overline{\Psi}_{2}\gamma_{5}\Psi
_{1}-\overline{\Psi}_{1}\gamma_{5}\Psi_{2}\right)  .
\end{equation}
(Eventually, the latter can be seen as a condensation of a tetraquark and/or a
glueball field, details in\ Refs. \cite{gallas}). The nucleon fields $N$ and
its chiral partner (associated to the resonance $N^{\ast}(1535)$) are obtained
as
\begin{align}
\Psi_{1}  &  =\frac{1}{\sqrt{{\small 2}\cosh{\small \delta}}}\left(
Ne^{\delta/2}+\gamma_{5}N^{\ast}e^{-\delta/2}\right)  \text{ ,}\\
\Psi_{2}  &  =\frac{1}{\sqrt{{\small 2}\cosh{\small \delta}}}\left(
\gamma_{5}Ne^{-\delta/2}-N^{\ast}e^{\delta/2}\right)  \text{ ,}%
\end{align}
where
\begin{equation}
\cosh\delta=\frac{m_{N}+m_{N^{\ast}}}{2m_{0}}\text{ .}%
\end{equation}
The value $m_{0}=460\pm136$ MeV was obtained by a fit to vacuum properties
\cite{gallas}.

We now write down a chirally invariant Lagrangian which describes the
interaction of $\tilde{G}$ with the baryon field $\Psi_{1}$ and $\Psi_{2}$
\begin{equation}
\mathcal{L}_{\tilde{G}\text{-baryons}}^{int}=ic_{\tilde{G}\Psi}\tilde
{G}\left(  \overline{\Psi}_{2}\Psi_{1}-\overline{\Psi}_{1}\Psi_{2}\right)
\text{ .} \label{intbar}%
\end{equation}
Thus, the fusion of a proton and an antiproton is described by $\mathcal{L}%
_{\tilde{G}\text{-baryons}}^{int},$ showing that it is not chirally
suppressed. Moreover, although the coupling constant $c_{\tilde{G}\Psi}$
cannot be determined, we can easily predict the ratio of the decay processes
$\Gamma_{\tilde{G}\rightarrow\overline{N}N}$ and $\Gamma_{\tilde{G}%
\rightarrow\overline{N^{\ast}}N+h.c.}$,
\begin{equation}
\frac{\Gamma_{\tilde{G}\rightarrow\overline{N}N}}{\Gamma_{\tilde{G}%
\rightarrow\overline{N}^{\ast}N+h.c.}}=1.94\text{ .} \label{ratio}%
\end{equation}

\section{Conclusion}

\qquad We have presented the chiral Lagrangians describing the interaction of
the pseudoscalar glueball with (pseudo)scalar mesons and baryons. In
particular, after the recall of mesonic effective Lagrangian of Eq.
(\ref{intbar}), and the corresponding results for the mesonic decays presented
in Ref. \cite{our} (see Table I.a and I.b), we have focused our attention on a
peculiar interference phenomenon taking place in the meson sector. The latter,
although subdominant, should be fully taken into account in future
studies.\ As a last step we have presented in Eq. (\ref{intbar}) the chiral
coupling of the pseudoscalar glueball with the nucleon and its chiral partner,
which describes the proton fusion process $\bar{p}p\rightarrow\tilde{G}$.
Finally, we have also made a prediction for the ratio of decays $\Gamma
_{\tilde{G}\rightarrow\overline{N}N}/\Gamma_{\tilde{G}\rightarrow\overline
{N}^{\ast}N+h.c.}=1.94$, which can be experimentally checked in the future.

\bigskip

\textbf{Acknowledgment}: The authors thank Dirk H. Rischke for useful
discussions. W.E.\ acknowledges support from DAAD and HGS-HIRe,
S.J.\ acknowledges support from H-QM and HGS-HIRe. F.G.\ thanks the Foundation
Polytechnical Society Frankfurt am Main for support through an Educator fellowship.

\end{document}